\documentclass[journal]{IEEEtran}
\usepackage{ifpdf}
\usepackage{graphicx}
\usepackage{amssymb}
\usepackage{amsmath}
\usepackage{cite}
\usepackage{epstopdf}
\usepackage{dsfont}
\usepackage[justification=centering]{caption}
\usepackage{color}
\usepackage{algorithm}
\usepackage{algorithmicx, algpseudocode}
\usepackage{etoolbox}
\usepackage{subcaption}

\newcommand{\mbf}[1]{\mathbf{#1}}

\newtheorem{theorem}{Theorem}

\makeatletter
\def\ScaleIfNeeded{%
\ifdim\Gin@nat@width>\linewidth \linewidth \else \Gin@nat@width \fi
} \makeatother

\begin{document}

\title{Sum Rate Maximization for IRS-assisted Uplink NOMA}

\author{\IEEEauthorblockN{Ming Zeng,
Xingwang Li, Gen Li, Wanming Hao and Octavia A. Dobre, \emph{Fellow, IEEE}
}
\thanks{
M. Zeng, G. Li and O. A. Dobre are with Memorial University, St. John's, NL A1B 3X9, Canada (e-mail: mzeng, lgen, odobre@mun.ca).

X. Li is with Henan Polytechnic University, Jiaozuo 454000, China (e-mail: lixingwangbupt@gmail.com). 

W. Hao is with Zhengzhou University, Zhengzhou 450001, China, (iewmhao@zzu.edu.cn). 

} 
}


\maketitle

\begin{abstract}
An intelligent reflecting surface (IRS) consists of a large number of low-cost reflecting elements, which can steer the incident signal collaboratively by passive beamforming. This way, IRS reconfigures the wireless environment to boost the system performance. 
In this paper, we consider an IRS-assisted uplink non-orthogonal multiple access (NOMA) system. The objective is to maximize the sum rate of all users under individual power constraint. The considered problem requires a joint power control at the users and beamforming design at the IRS, and is non-convex. To handle it, semidefinite relaxation is employed, which provides a near-optimal solution. Presented numerical results show that the proposed NOMA-based scheme achieves a larger sum rate than orthogonal multiple access (OMA)-based one. Moreover, the impact of the number of reflecting elements on the sum rate is revealed.   




\end{abstract}

\begin{IEEEkeywords}
Intelligent reflecting surface (IRS), non-orthogonal
multiple access (NOMA), sum rate, uplink.
\end{IEEEkeywords}
\IEEEpeerreviewmaketitle


\section{Introduction}
Recently, the intelligent reflecting surface (IRS) technology has received significant attention for beyond 5G systems \cite{Basar_Access19}. An IRS is a planar array consisting of multiple low-cost reflecting elements. Each element can steer the incident signal through passive beamforming, and thus, IRS can reconfigure the wireless environment to facilitate information transmission \cite{Guo_Arxiv19, Nadeem_Arxiv19, Wu_TWC19, Huang_TWC19}. 
The authors in \cite{Guo_Arxiv19} considered the weighted sum rate maximization when a direct link exists between the base station (BS) and users. The formulated problem required to consider both active and passive beamforming, and was handled using alternating optimization. The authors in \cite{ Nadeem_Arxiv19} aimed to maximize the minimum rate of all users to take into account system fairness. Different from \cite{Guo_Arxiv19}, the direct link between the BS and users was assumed blocked. The authors in \cite{ Wu_TWC19} considered the power minimization at the BS subject to users' individual minimum rate constraints. The single user case was first considered to reveal insight. Then, the multi-user case was handled as an extension. Additionally, the authors in \cite{ Huang_TWC19} studied the system energy efficiency maximization. Two computationally affordable approaches were proposed, capitalizing on alternating maximization, gradient descent search, and sequential fractional programming. 

Note that the works \cite{Guo_Arxiv19, Nadeem_Arxiv19, Wu_TWC19, Huang_TWC19} are all based on orthogonal multiple access (OMA). Recently, 
a few works started to investigate the application of IRS to non-orthogonal multiple access (NOMA) networks to further boost the system performance \cite{Ding_CL20, Yang_Arxiv20, Hou_Arxiv20}. In \cite{Ding_CL20}, the authors proposed a simple design of IRS-assisted NOMA transmission to ensure that more users are served on each orthogonal spatial direction when compared with spatial division multiple access. In \cite{Yang_Arxiv20}, the authors studied the joint optimization of the active beamforming vectors at the
BS and the passive beamforming at the IRS to maximize the minimum target decoding signal-to-interference-plus-noise-ratio (SINR) of all users. In \cite{Hou_Arxiv20}, the authors considered a novel IRS-aided NOMA network, where a priority-oriented design was proposed to enhance the sum rate. The impact of the proposed design on the system's outage probability (OP), ergodic rate, and sum rate was analyzed.

Unlike the above mentioned works \cite{Ding_CL20, Yang_Arxiv20, Hou_Arxiv20}, which focus on downlink, we consider an IRS-assisted uplink NOMA system. 
To the best of our knowledge, this is the first work to consider IRS-assisted NOMA under uplink. The objective is to maximize the sum rate of all users under individual power constraints. The formulated problem requires to jointly optimize the power at the users and phase shifts at the IRS, and is non-convex. In particular, the main challenge lies in the element-wise constant modulus constraint at the IRS. A suboptimal solution based on semidefinite relaxation (SDR) is proposed. Numerical results are presented, showing that the proposed solution achieves near-optimal performance, and outperforms the OMA-based counterpart in terms of sum rate.  


\section{System Model and Problem Formulation}
\subsection{System Model}
As shown in Fig.~1, we consider an uplink system, where $K$ users aim to communicate with the BS. Both users and BS are equipped with a single antenna. It is assumed that no direct link exists between the users and BS due to unfavorable propagation conditions. Therefore, this communication takes place via an IRS with $N$ reflecting elements deployed on the facade of a building located in the proximity of both communication ends.


The signal received at the BS is given by 
\begin{align} \label{y}
y = \sum_{k=1}^K \mbf{h}_{\rm{BS}}^H \mbf{\Phi} \mbf{h}_{k} \sqrt{P_k} s_k + n,
\end{align}
where $s_k$ denotes the signal from user $k$ and is of unit power, i.e., $\mathbb{E}[|s_{k}|^2]=1$, 
$k \in \{1, \cdots, K\}$,
with $\mathbb{E}$ being the expectation operation. 
$P_k$ represents the corresponding transmit power, satisfying $P_k \leq P_k^{\max}$, with $P_k^{\max}$ being the maximum transmit power. $\mbf{h}_{k} \in \mathbb{C}^{N \times 1}$ denotes the channel vector between user $k$ and IRS, while $\mbf{h}_{\rm{BS}} \in \mathbb{C}^{N \times 1}$ represents that between the IRS and BS. $\mbf{\Phi} = {\rm{diag}}[\phi_1, \phi_2, \cdots, \phi_N]$ is a diagonal matrix accounting for the effective phase shifts from all IRS reflecting elements, satisfying $|\phi_i| = 1$, $\forall i \in \{1, \cdots, N\}$. This constant modulus constraint is because IRS simply reflects the received signal and cannot amplify it.
$n \sim \mathcal{CN} (0, \sigma^2)$ denotes the additive white Gaussian noise at the BS. 

\begin{figure}
\centering
\includegraphics[width=0.45\textwidth]{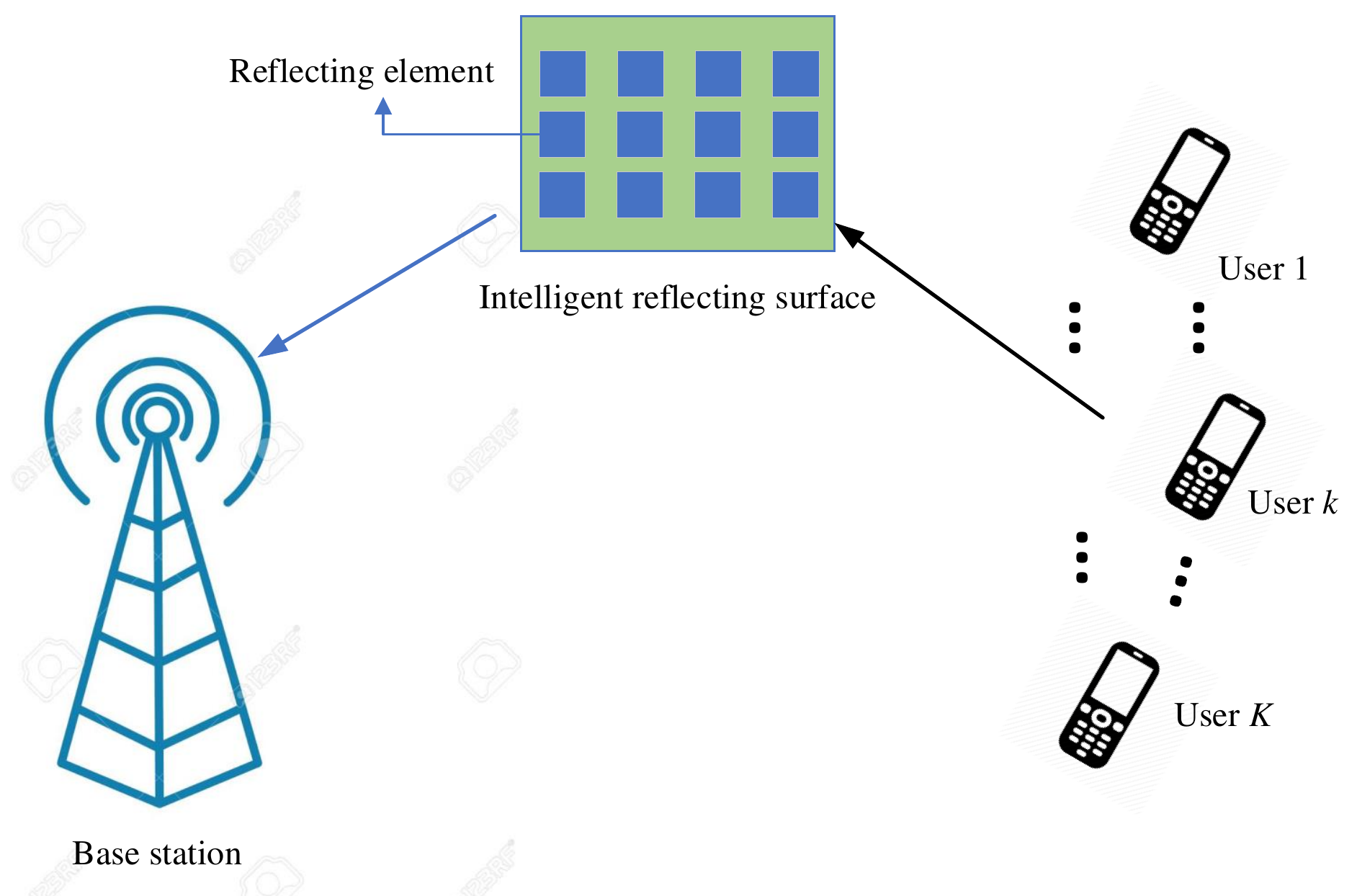}
\caption{IRS-assisted uplink NOMA.}
\end{figure}

From \eqref{y} it is clear that user $k$ receives other users' interference. To mitigate this interference, the BS performs successive interference cancellation (SIC) using NOMA.
For uplink NOMA, the users with better channel conditions are often decoded earlier. Here, the effective channel for user $k$ is $\mbf{h}_{\rm{BS}}^H \mbf{\Phi} \mbf{h}_{k} $, which depends on the unknown parameter $\mbf{\Phi}$. Therefore, we cannot order the users using the effective channels. To address this issue, we remove the unknown parameter $\mbf{\Phi}$ from the effective channel, and consider the simplified effective channel $\mbf{h}_{\rm{BS}}^H \mbf{h}_{k}$. Without loss of generality, we further assume that the users are arranged in a descending order of the simplified effective channels, namely
\begin{equation}
|\mbf{h}_{\rm{BS}}^H \mbf{h}_{1}| \geq \cdots \geq |\mbf{h}_{\rm{BS}}^H \mbf{h}_{K}|.
\end{equation} 

According to NOMA protocol, the SINR of user $k$ can be expressed as \cite{Ming_IoT19}
\begin{equation}
\gamma_k = \frac{|\mbf{h}_{\rm{BS}}^H \mbf{\Phi} \mbf{h}_{k}|^2 P_k}{\sum_{i=k+1}^K |\mbf{h}_{\rm{BS}}^H \mbf{\Phi} \mbf{h}_{i}|^2 P_i + \sigma^2},
\end{equation}
where $\sum_{i=k+1}^K |\mbf{h}_{\rm{BS}}^H \mbf{\Phi} \mbf{h}_{i}|^2 P_i=0$ when $k=K$. 

Then, the achievable data rate of user $k$ is given by
\begin{equation}
R_k = \log_2 (1+ \gamma_k).
\end{equation}

\subsection{Problem Formulation}
In this paper, we aim to maximize the sum rate of the users through appropriate passive beamforming at the IRS and power control at the users. 
The considered problem can be formulated as follows:
\begin{subequations}\label{P1}
\begin{align} 
 \underset{\mbf{P}, \mbf{\Phi}}{\max} &~ \sum_{k=1}^K \log_2 (1+ \gamma_k) \\
\text{s.t.}
& ~ P_k \leq P_k^{\max}, \forall k \in \{1, \cdots, K\} \\
& ~ |\phi_i| = 1, \forall i \in \{1, \cdots, N\},
\end{align}
\end{subequations}
where $\mbf{P} = [P_1, \cdots, P_K]$ denotes the transmit power vector. 
 
\section{Proposed Solution}
Problem \eqref{P1} is non-convex due to the non-convex objective function and constraint (\ref{P1}c). 
To handle it, let us first look at the objective function, which can be re-expressed as 
\begin{align} \label{sum rate} 
R_{\rm{sum}} 
&= \sum_{k=1}^K \log_2 \left(1+ \frac{|\mbf{h}_{\rm{BS}}^H \mbf{\Phi} \mbf{h}_{k}|^2 P_k}{\sum_{i=k+1}^K |\mbf{h}_{\rm{BS}}^H \mbf{\Phi} \mbf{h}_{i}|^2 P_i + \sigma^2} \right) \\ \nonumber
&= \log_2 \left(1+ \frac{\sum_{k=1}^K |\mbf{h}_{\rm{BS}}^H \mbf{\Phi} \mbf{h}_{k}|^2 P_k}{ \sigma^2} \right),
\end{align}
where the last equality holds since the terms inside the brackets in the sum rate expression forms a telescoping product. From \eqref{sum rate}, the sum rate is a function of $\sum_{k=1}^K |\mbf{h}_{\rm{BS}}^H \mbf{\Phi} \mbf{h}_{k}|^2 P_k $, which is independent of user ordering. As a result, it can be claimed that the sum rate for IRS-assisted NOMA is independent of the decoding order. 

Moreover, regarding the optimal transmit power, we have the following theorem:
\begin{theorem} \label{theorem 1}
Regardless of the beamforming matrix at the IRS, each user simply transmits at full power to maximize the sum rate. 
\end{theorem}
\begin{IEEEproof}
Under any given value of $\mbf{\Phi}$,  the sum rate is monotonic increasing with $P_k$ according to \eqref{sum rate}. Therefore, each user should transmit at full power to maximize the sum rate.
\end{IEEEproof}

Based on Theorem~\ref{theorem 1}, problem \eqref{P1} can be simplified as
\begin{subequations}\label{P2}
\begin{align} 
 \underset{\mbf{\Phi}}{\max} &~ \log_2 \left(1+ \frac{\sum_{k=1}^K |\mbf{h}_{\rm{BS}}^H \mbf{\Phi} \mbf{h}_{k}|^2 P_k^{\max}}{ \sigma^2} \right) \\
\text{s.t.}
& ~ |\phi_i| = 1, \forall i \in \{1, \cdots, N\},
\end{align}
\end{subequations}
where the only variable is $\mbf{\Phi}$. 

Since the $\log$ function is a monotonic increasing one, maximizing $\log_2 \left(1+ \frac{\sum_{k=1}^K |\mbf{h}_{\rm{BS}}^H \mbf{\Phi} \mbf{h}_{k}|^2 P_k^{\max}}{ \sigma^2} \right)$ is equivalent to maximizing $ \sum_{k=1}^K |\mbf{h}_{\rm{BS}}^H \mbf{\Phi} \mbf{h}_{k}|^2 P_k^{\max}$. 
To handle $|\mbf{h}_{\rm{BS}}^H \mbf{\Phi} \mbf{h}_{k}|^2$ more easily, we first re-arrange the diagonal matrix $\mbf{\Phi}$ into a vector $\mbf{w} \in \mathbb{C}^{N \times 1}$, with element $w_i = \phi_i^H$, $\forall i \in \{1, \cdots, N\}$. It is clear that $\mbf{w}$ contains all the information of $\mbf{\Phi}$. Then, we introduce an auxiliary vector $\widehat{\mbf{h}}_k = \mbf{h}_{\rm{BS}} \circ \mbf{h}_{k}$, where the operation $\circ $ represents the Hadamard product. Accordingly, it can be easily verified that the following equality holds: 
\begin{equation}
|\mbf{h}_{\rm{BS}}^H \mbf{\Phi} \mbf{h}_{k}|^2= |\mbf{w}^H \widehat{\mbf{h}}_k|^2. 
\end{equation}

Then, problem \eqref{P2} can be re-expressed as
\begin{subequations}\label{P3}
\begin{align} 
 \underset{\mbf{w}}{\max} &~  \sum_{k=1}^K |\mbf{w}^H \widehat{\mbf{h}}_k|^2 P_k^{\max} \\
\text{s.t.}
& ~ |w_i| = 1, \forall i \in \{1, \cdots, N\}.
\end{align}
\end{subequations}

The difficulty of solving problem \eqref{P3} mainly comes from the non-convex objective function. To address this issue, we further reformulate $\sum_{k=1}^K |\mbf{w}^H \widehat{\mbf{h}}_k|^2 P_k^{\max}$ as follows:
\begin{subequations}
\begin{align}
\sum_{k=1}^K |\mbf{w}^H \widehat{\mbf{h}}_k|^2 P_k^{\max} & = \sum_{k=1}^K \mbf{w}^H \widehat{\mbf{h}}_k \widehat{\mbf{h}}_k^H \mbf{w} P_k^{\max} \\
& = \mbf{w}^H \sum_{k=1}^K P_k^{\max}  \widehat{\mbf{h}}_k \widehat{\mbf{h}}_k^H \mbf{w} .
\end{align}
\end{subequations}

Now we introduce an auxiliary matrix $\mbf{H}=\sum_{k=1}^K P_k^{\max}  \widehat{\mbf{h}}_k \widehat{\mbf{h}}_k^H$. It can be easily verified that $\mbf{H}$ is a positive semi-definite matrix. Meanwhile, problem \eqref{P3} can be re-expressed as
\begin{subequations}\label{P4}
\begin{align} 
 \underset{\mbf{w}}{\max} &~  \mbf{w}^H \mbf{H} \mbf{w} \\
\text{s.t.}
& ~ |w_i| = 1, \forall i \in \{1, \cdots, N\}.
\end{align}
\end{subequations}

The objective function (\ref{P4}a) is still non-convex, since we aim to maximize a quadratic function with a positive semi-definite matrix. Meanwhile, constraint (\ref{P4}b) is also non-convex. Nonetheless, the following theorem gives the upper bound on the optimal solution:

\begin{theorem} \label{theorem2}
The optimal solution for \eqref{P4} is upper bounded by $\lambda N^2$, where $\lambda$ is the maximum eigenvalue of $\mbf{H}$. 
\end{theorem}

\begin{IEEEproof}
We first relax the element-wise constraint (\ref{P4}b) to a sphere constraint $|\mbf{w}|_2 = N$. Then, \eqref{P4} can be reformulated as
\begin{subequations}\label{P5}
\begin{align} 
 \underset{\mbf{w}}{\max} &~  \mbf{w}^H \mbf{H} \mbf{w} \\
\text{s.t.}
& ~ |\mbf{w}|_2 = N.
\end{align}
\end{subequations}

It has been shown in \cite{Hager_01} that the maximum value for \eqref{P5} is obtained when $\mbf{w} = N \mbf{x}$, where $\mbf{x}$ is the normalized eigenvector of $\mbf{H}$ with regard to the largest eigenvalue $\lambda$. Moreover, the optimal solution is given by $ \mbf{w}^H \mbf{H} \mbf{w}=\lambda N^2  \mbf{x}^H  \mbf{x}=\lambda N^2 $. Since we expand the feasible region by relaxing (\ref{P4}b) to (\ref{P5}b), $\lambda N^2$ is the upper bound. 
\end{IEEEproof}

According to Theorem~\ref{theorem2}, the SINR of the sum rate can have the potential of growing with the square of the number of elements at the IRS. 

We now try to find an appropriate feasible solution to \eqref{P4}. It is first observed that
\begin{equation}
\mbf{w}^H \mbf{H} \mbf{w} = \rm{Tr} (   \mbf{w}^H \mbf{H} \mbf{w} ) = \rm{Tr} (    \mbf{H} \mbf{w}\mbf{w}^H ), 
\end{equation}
where $\rm{Tr}(\mbf{X})$ represents the trace of $\mbf{X}$. 

By introducing a new variable $ \mbf{W} =\mbf{w}\mbf{w}^H $, \eqref{P4} can be equivalently transformed into the following problem:
\begin{subequations}\label{P6}
\begin{align} 
 \underset{\mbf{W}}{\max} &~  \rm{Tr} (  \mbf{H} \mbf{W} ) \\
\text{s.t.}
& ~ {\rm{Tr}} (  \mbf{B}_i \mbf{W} ) = 1, \forall i \in \{1, \cdots, N\} \\
& ~ \mbf{W} \succeq \mbf{0}, \\
& ~ {\rm{rank}} (\mbf{W})=1,
\end{align}
\end{subequations}
where (\ref{P6}c) means that $\mbf{W}$ is a symmetric positive semidefinite matrix. (\ref{P6}d) limits the rank of $\mbf{W}$ to 1, since $ \mbf{W} =\mbf{w}\mbf{w}^H $. 
$\mbf{B}_i  \in \mathbb{R}^{N \times N}$ has only one non-zero element, which is located at the $(i,i)$-th position and equals to 1. That is, $\mbf{B}_i(i,i)=1$ and the rest elements of $\mbf{B}_i$ are all zeros. Therefore, ${\rm{Tr}} (  \mbf{B}_i \mbf{W} )=\mbf{W}(i,i) = w_i^2=1$, which is the same as $|w_i| = 1$.    

Problem \eqref{P6} is a semidefinite programming, and the only non-convex constraint is the rank constraint (\ref{P6}d). By dropping (\ref{P6}d), we can obtain the following relaxed version of \eqref{P4}
\begin{subequations}\label{P7}
\begin{align} 
 \underset{\mbf{W}}{\max} &~  \rm{Tr} (  \mbf{H} \mbf{W} ) \\
\text{s.t.}
& ~ {\rm{Tr}} (  \mbf{B}_i \mbf{W} ) = 1, \forall i \in \{1, \cdots, N\} \\
& ~ \mbf{W} \succeq \mbf{0}.
\end{align}
\end{subequations}

Problem \eqref{P7} is known as an SDR of \eqref{P4}, and can be solved efficiently by readily available software packages, e.g., CVX. Denote the obtained optimal solution from solving \eqref{P7} by $\mbf{W}^{\star}$.
Now the problem lies in how to convert $\mbf{W}^{\star}$ to a feasible solution of \eqref{P4}. If ${\rm{rank}} (\mbf{W}^{\star})=1$ holds, then we can write $ \mbf{W}^{\star} =\mbf{w}^{\star} {\mbf{w}^{\star}}^H $, and $\mbf{w}^{\star}$ will be a feasible and indeed optimal solution to \eqref{P4}. On the other hand, if ${\rm{rank}} (\mbf{W}^{\star}) >1$, we then need to extract a feasible solution $\widehat{\mbf{w}} $ for \eqref{P4} from $\mbf{W}^{\star} $. An effective way to do this is to select the eigenvector of $\mbf{W}^{\star} $ that corresponds to its largest eigenvalue, since this provides the best rank-one approximation \cite{Luo_SPM10}. Denote the largest eigenvalue of $\mbf{W}^{\star} $ and its associated eigenvector by $\lambda_1$ and $\mbf{q}_1 $, respectively. Then, $\widetilde{\mbf{w}}= \sqrt{\lambda_1} \mbf{q}_1 $ may be used as our candidate solution to \eqref{P4}, provided that it is feasible. 
If $\widetilde{\mbf{w}}$ is infeasible, we need to map $\widetilde{\mbf{w}}$ to a ``nearby'' feasible solution $\widehat{\mbf{w}} $. This can be done by normalizing each element of $\widetilde{\mbf{w}}$ to 1, i.e., $\widehat{{w}}_i= \frac{\widetilde{{w}}_i}{|\widetilde{{w}}_i|}, \forall i \in \{1, \cdots, N\}  $.

\section{Baseline Scheme: OMA}
For OMA, it is assumed that the degrees-of-freedom (time/frequency) are equally divided among the $K$ users.
In this case, the passive beamforming of the IRS and power control at the user can be performed separately for each user. Without loss of generality, we consider user $k$ here, whose achievable rate is given by
\begin{equation}
R_k^{\rm{OMA}} =\frac{1}{K} \log_2 \left(1+ \frac{ K |\mbf{h}_{\rm{BS}}^H \mbf{\Phi} \mbf{h}_{k}|^2 P_k}{ \sigma^2} \right). 
\end{equation}

After a similar procedure as for NOMA, we have the following problem corresponding to \eqref{P3} in NOMA:
\begin{subequations} \label{OMA}
\begin{align} 
 \underset{\mbf{w}}{\max} &~  |\mbf{w}^H \widehat{\mbf{h}}_k|^2 P_k^{\max} \\
\text{s.t.}
& ~ |w_i| = 1, \forall i \in \{1, \cdots, N\}.
\end{align}
\end{subequations}

It can be easily verified that the optimal beamforming satisfies $w_i = \frac{ \widehat{\mbf{h}}_k (i)}{|\widehat{\mbf{h}}_k (i)|}, \forall i \in \{1, \cdots, N\}  $.


\begin{figure}
\centering
\includegraphics[width=0.48\textwidth]{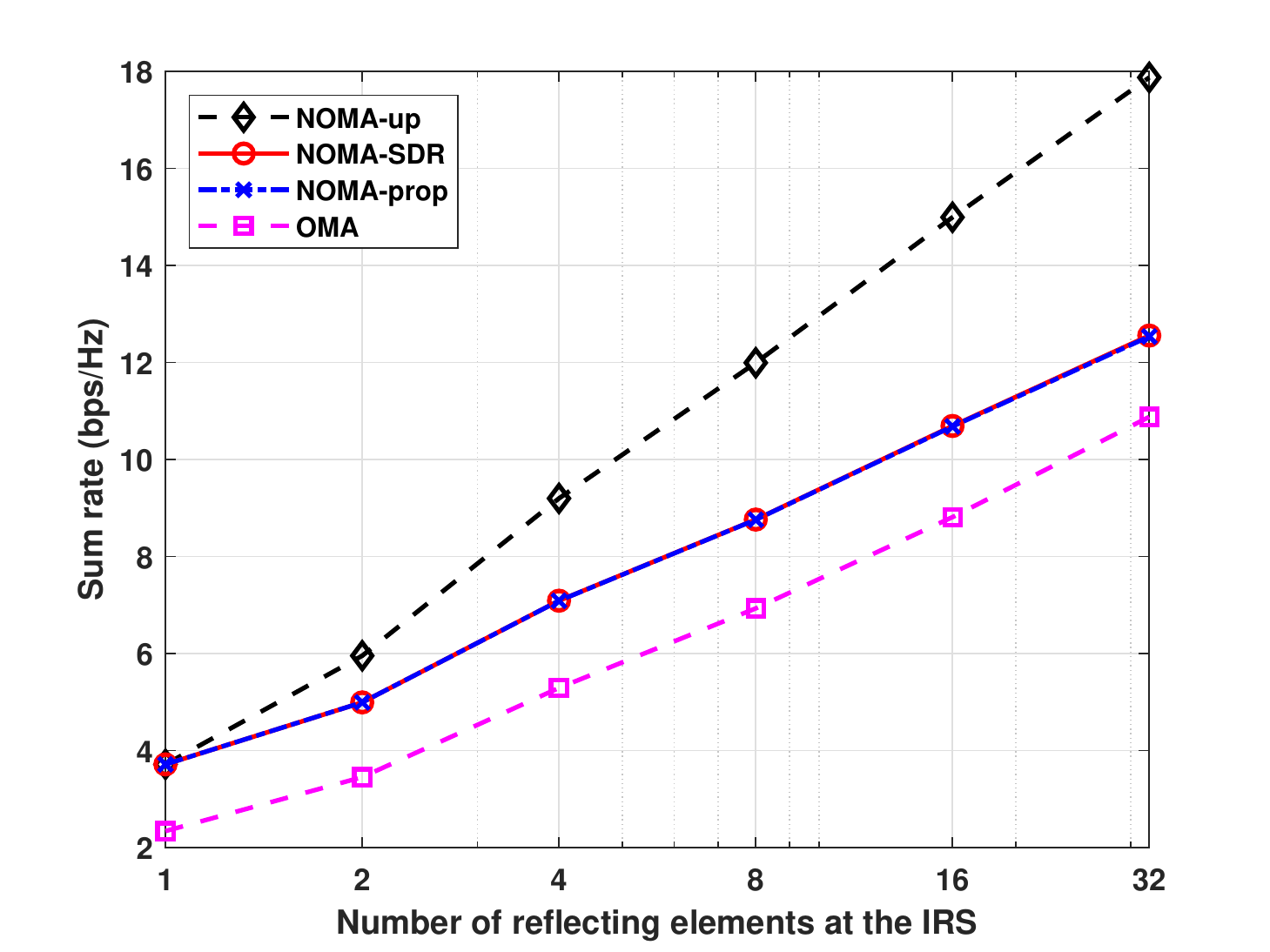}
\caption{Sum rates versus the number of reflecting element at the IRS, $N$ under different schemes.}
\end{figure}
\section{Simulation Results}
In this section, numerical results are presented to compare the performance of the two considered schemes. The default simulation parameters are as follows \cite{Guo_Arxiv19, Wu_TWC19}: the number of users and reflecting elements at the IRS is 3 and 16, respectively. The distances between the BS and IRS and that between the IRS and users are generated randomly uniformly within 50 and 200 m, respectively. The channel between the BS and IRS is assumed to follow the free-space propagation model. In contrast, the large-scale pathloss model of the channels between the IRS and users follows $30+28 \log_{10}(d)$, where $d$ is the distance in m. Moreover, Rayleigh fading is used for small-scale fading. The bandwidth is $B = 1$ MHz, while the noise power spectral density is $N_0=174$ dBm/Hz. All results are averaged over $10^3$ random trials. 

Figure 1 plots the sum rate versus the number of reflecting elements at the IRS under different schemes. More specifically, {NOMA-SDR} denotes the result of NOMA from solving \eqref{P7}, while {NOMA-prop} represents the proposed feasible solution from converting the {NOMA-SDR} solution.  {NOMA-up} denotes the upper bound of NOMA from Theorem~2, whereas {OMA} is the solution from \eqref{OMA}. 
It can be seen that the sum rates for all considered schemes grow linearly with $\log(N)$, which illustrates the effectiveness of having more reflecting elements at the IRS. By comparing the three NOMA schemes, it is clear that {NOMA-SDR} provides a much tighter upper bound than {NOMA-up}. Meanwhile, {NOMA-SDR} and {NOMA-prop} perform close to each other,
which indicates that {NOMA-prop} is close to optimal. In particular, it is found that the obtained solutions $\mbf{W}^{\star}$ from {NOMA-SDR} has exactly one large eigenvalue (close to $N$) and all other eigenvalues are almost zero.\footnote{The sum of the eigenvalues from $\mbf{W}^{\star}$ equals to $N$, since ${\rm{Tr}}(\mbf{W})=\sum_{i=1}^N w_i^2=N$.} Additionally, it is clear that {NOMA-prop} dominates {OMA} in terms of sum rate for any given value of $N$, verifying the superiority of NOMA over OMA. 
 
Figure 2 further shows the sum rates versus the number of users under the considered schemes. It can be seen that the sum rates under all considered schemes grow with the number of users $K$. However, the increase declines under a larger value of $K$. This behavior can be explained by the concavity of the $\log$ function from the sum rate. In addition, as expected, {NOMA-prop} obtains the same sum rate as {OMA} when $K=1$. When $K>1$, {NOMA-prop} outperforms {OMA} in terms of sum rate and the gap increases with $K$. This increase may be because in NOMA, a multiplexing gain is available, since all users are served by the IRS simultaneously. In contrast, only a power gain is available for OMA, since users are served using orthogonal resources. The increase may also come from the interference cancellation of SIC in NOMA.

\begin{figure}
\centering
\includegraphics[width=0.48\textwidth]{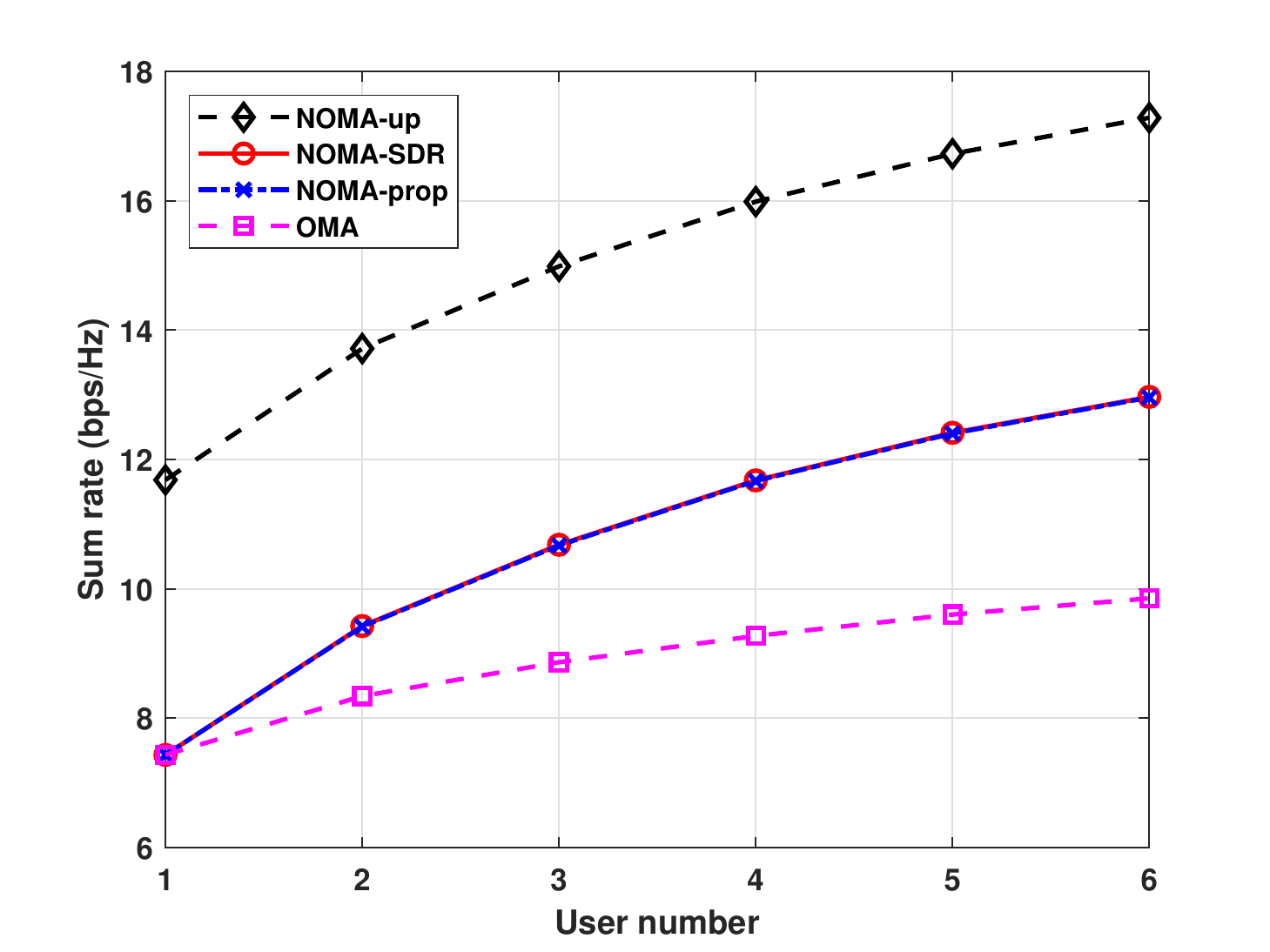}
\caption{Sum rates versus the number of users, $K$ under different schemes.}
\end{figure}

\section{Conclusion}
In this paper, we considered the sum rate maximization problem for an IRS-assisted uplink NOMA system. The formulated problem involves a joint optimization of the power at the users and the passive beamforming at the IRS, and is non-convex. In particular, the main challenge lies in the element-wise constant modulus constraint at the IRS. To address it, we transformed it into an SDR, and solved it accordingly. The proposed NOMA scheme was compared with its OMA-based counterpart, for which a closed-form solution was derived. The superiority of NOMA over OMA was fully validated by the numerical results. Moreover, it was shown that the sum rates for both NOMA and OMA grow linearly with $\log(N)$, illustrating the effectiveness of employing a large number of reflecting elements at the IRS.

\bibliographystyle{IEEEtran}
\bibliography{IEEEabrv,conf_short,jour_short,mybibfile}

\end{document}